\documentclass[11pt]{article}
\usepackage{amssymb}
\usepackage{amsmath}

\begin{document}

\title{Comment -- Practical Data Protection}

\author{Manik Lal Das\\Dhirubhai Ambani Institute of Information and Communication Technology\\Gandhinagar - 382007, India.\\
Email: maniklal.das@ieee.org}

\date{}
\maketitle

\begin{abstract}
Recently, Rawat and Saxena proposed a method for protecting data
using ``Disclaimer Statement''. This paper presents some issues
and several flaws in their proposal.
\end{abstract}

\section{Data Protection using ``Disclaimer Statement''}
Recently, Rawat and Saxena \cite{rwa08} proposed a method for
securing data using ``disclaimer statement'' in email. Typically,
the disclaimer statement is appended at the end of email. But,
Rawat and Saxena intended to put the disclaimer statement at the
beginning of the email. They claim that their method achieves the
following characteristics:
\newcounter{1}
\begin{list}{-}
{\usecounter{1}} \item the method provides very tight
confidentiality and could be extended towards creative disclaimers
based on security requirements and level. \item As subject of the
email may give some clue to the receiver, no subject on
``Subject'' field of the email. In fact, they suggested to write
``subject'' after the disclaimer. \item the ``disclaimer'' method
makes all current encryption methods obsolete.
\end{list}
We refer interested reader to \cite{rwa08} for a quick review of
their proposal. In next section, we show that none of the their
claims provides any data protection mechanisms.

\section{Comments and Remarks}
I cannot see any data protection method in Rawat and Saxena's
method \cite{rwa08}. They neither provide any methods for data
confidentiality nor safeguard data privacy. The picture of their
proposal \cite{rwa08} would be pretty clear from the following
comments, which one could easily observe these flaws from their
proposal.

\newcounter{2}
\begin{list}{-}
{\usecounter{2}} \item The basic purpose of email is to
communicate message in between sender and receiver. Although some
unsolicited junk emails are not new these days; however, one may
adopt some strategies to safeguard/minimize mail server from
receiving unsolicited emails.\\
The basic purpose of ``disclaimer statement'' attached to email is
to defend the sender/organization's intention (accidental or
intentional) against data abusing in a court of law, if such types
of circumstances occur. And that's why ``disclaimer statement'' is
appended in email and it works perfectly to meet such intention on
top of basic security requirement of email messaging.\\
Rawat and Saxena advocate to put ``disclaimer statement'' at the
beginning, even before displaying the ``subject'' of the mail.
Well, it wouldn't cost if we shift the ``disclaimer statement''
from bottom to top, but, this is against some basic ethics of
email messaging. The purpose of disclaimer statement is to avoid
abuse of information, if any. Certain basic questions arise as
follows: Why does the receiver read disclaimer statement before
reading the actual message/subject of the email? How does s/he
come to know whether the message intends to her/him. Is it
possible to smell the taste of the email's message just by seeing
sender email-id? How does ``disclaimer statement'' prevent the
receiver not to read the body of message? Where is data
protection?...\\
Consequently, ``disclaimer statement'' should be appended at the
end of the message or after providing some hints about the message
or sender's intention to the receiver. \item If no ``subject'' and
``no message information'' then why ``disclaimer statement''? In
real-world scenarios, ``subject'' not only provides
intention/requirement, but also reduces the cost of the task. I
think ``subject'' information on the ``subject line'' of email
provides huge information to the receiver about (un)importance of
the email. One may delete the email on seeing the subject line,
without going through the body or ``disclaimer statement''. \item
A funny argument provided by Rawat and Saxena is that they have a
strong security analysis of their method received from the
anonymous reviewers, but because of the ``disclaimer statement''
of the received message they cannot present that security
analysis. How come ``disclaimer statement'' prevents it? If the
reviewers asked them not to present then fine. Instead, they
should either delete the reviewers security analysis part or
present it with proper acknowledgement. Let us take a simple
example: suppose the Red server receives an email from Bluechip
(who is known and trusted) with strong disclaimer statement,
stating that she has suddenly found an algorithm which can factor
large integer in polynomial time. The Red server then announces
the news and potential threat warning through a blog, which
luckily got attraction of crypto community. On seeing the news,
will the community believe it blindly or check in detail? Will the
community throw away all systems, where security relies on integer
factorization problem? \item Rawat and Saxena think their method
makes all real-world encryption algorithms obsolete. Where is the
confidentiality of data while using disclaimer method? One can
read the data when s/he has received email, even though the email
contains strong or weak ``disclaimer statement'' at the beginning
or end. Are they aiming at ``zero key'' data confidentiality? If
so, that would certainly create a new era of cryptography. Of
course, this can be true if (and only if) only Alice and Bob live
in this universe. However, if by chance with a non-negligible
probability, Eve comes into the picture as a form of Unidentified
Flying Object, then achieving ``zero key'' data confidentiality is
near to impossible task. Further, email messaging is not a secure
mechanism to protect data. How does disclaimer statement protect
border, military applications, healthcare, trading, ...??
\end{list}

\section{Conclusion}
I do not see any data protection mechanism of Rawat and Saxena's
method; however, they provide a good pointer which could be used
for email protection by some additional measures such as
appropriateness of disclaimer statement's position and phrases on
top of other security properties of email messaging. Nevertheless,
while writing this note, ``disclaimer statement'' has not found
enough potentials to stop other encryption methods, consequently,
making statement ``...this new method makes all encryption method
obsolete'' in \cite{rwa08} is a baseless statement/claim. I guess,
disclaimer method may provide some sort of data protection in the
future to come by some additional measures, and lets see how much
we can utilize this idea for protecting (practical) data.

\end{document}